\documentclass[aps,preprint,superscriptaddress]{revtex4-1}
\usepackage{graphicx}  
\usepackage{dcolumn}   
\usepackage{bm}        
\usepackage{amssymb}   
\usepackage{amsmath}
\usepackage{xcolor}

\begin{document}

\title{Nonlinear spatiotemporal control of laser intensity}

\author{Tanner T. Simpson}
\email{tsim@lle.rochester.edu} 
\affiliation{
Laboratory for Laser Energetics, University of Rochester, Rochester, NY 14623, USA}
\author{Dillon Ramsey}
\affiliation{
Laboratory for Laser Energetics, University of Rochester, Rochester, NY 14623, USA} 

\author{Philip Franke}
\affiliation{
Laboratory for Laser Energetics, University of Rochester, Rochester, NY 14623, USA}
\author{Navid Vafaei-Najafabadi}
\affiliation{
Stony Brook University, Department of Physics and Astronomy, Stony Brook, NY 11794, USA}
\author{David Turnbull}
\affiliation{
Laboratory for Laser Energetics, University of Rochester, Rochester, NY 14623, USA}
\author{Dustin H. Froula}
\affiliation{
Laboratory for Laser Energetics, University of Rochester, Rochester, NY 14623, USA}
\author{John P. Palastro}
\email{jpal@lle.rochester.edu} 
\affiliation{
Laboratory for Laser Energetics, University of Rochester, Rochester, NY 14623, USA}

\date{\today}

\email{tsim@lle.rochester.edu} 
\email{jpal@lle.rochester.edu} 



\begin{abstract}
Spatiotemporal control over the intensity of a laser pulse has the potential to enable or revolutionize a wide range of laser-based applications that currently suffer from the poor flexibility offered by conventional optics. Specifically, these optics limit the region of high intensity to the Rayleigh range and provide little to no control over the trajectory of the peak intensity. Here, we introduce a nonlinear technique for spatiotemporal control, the ``self-flying focus,'' that produces an arbitrary trajectory intensity peak that can be sustained for distances comparable to the focal length. The technique combines temporal pulse shaping and the inherent nonlinearity of a medium to customize the time and location at which each temporal slice within the pulse comes to its focus. As an example of its utility, simulations show that the self-flying focus can form a highly uniform, meter-scale plasma suitable for advanced plasma-based accelerators.
\end{abstract}

\maketitle

\section{Introduction}
A wide range of laser-based applications share two requirements: (1) that the driving laser pulse maintain a high intensity over an extended distance, and (2) that the velocity of the peak intensity conform to some underlying process. Examples from across the fields of optics and plasma physics, such as THz \cite{d2007conical,kim2008coherent, hebling2002velocity} and high-harmonic generation \cite{lewenstein1994theory, rundquist1998phase, popmintchev2012bright}, photon acceleration \cite{dias1997experimental,mendonca2000theory}, laser wakefield and vacuum electron acceleration \cite{tajima1979laser,esarey2009physics}, Raman amplification \cite{malkin1999fast, trines2011simulations}, and plasma channel \cite{durfee1995development, milchberg1996development} or filament formation \cite{braun1995self,couairon2007femtosecond}  illustrate the ubiquity of these requirements. In THz generation by optical rectification, for instance, phase matching requires that the velocity of an intensity peak equal the phase velocity of the THz radiation, while maximizing the efficiency requires sustaining that intensity over the THz absorption length \cite{boyd2019nonlinear}. With respect to laser-wakefield acceleration, the velocity of the intensity peak determines the phase velocity of the driven plasma wave and, as a result, the distance it takes a relativistic electron to outrun the accelerating phase of the wave, i.e., the dephasing length \cite{esarey2009physics}. This length and the distance over which the pulse maintains a high intensity can limit the length of the accelerator and therefore the maximum electron energy.

By providing unprecedented control over the trajectory of an intensity peak and the distance over which it can be sustained, spatiotemporal pulse shaping promises to expand the design space for these applications \cite{froula2018spatiotemporal, sainte2017controlling,li2020velocity, howard2019photon, palastro2020dephasingless, caizergues2020phase, ramsey2020vacuum, turnbull2018raman, turnbull2018ionization}. Conventional optics rely on spatial shaping alone, e.g., through refraction or diffraction, which severely constrains their flexibility. While axicons and waveguides can extend the range of high intensity beyond the Rayleigh range of an ideal lens, none of these elements provide independent control over the velocity. One of the earliest successes of spatiotemporal pulse shaping was the use of a tilted pulse front, formed by a misaligned grating pair, to phase-match THz generation in a crystal \cite{hebling2002velocity}. This technique, however, lacks cylindrical symmetry and does not extend the range of high intensity. More recent spatiotemporal pulse shaping schemes, i.e., ``flying focus'' techniques, can produce cylindrically symmetric intensity peaks that propagate at any velocity over distances much longer than a Rayleigh range. As an example, the chromatic aberration of a diffractive optic and a chirp can be used to control the location and time at which each temporal slice within a pulse comes to its focus, respectively \cite{froula2018spatiotemporal, sainte2017controlling}. The velocity of the intensity peak in this scheme can be readily adjusted by changing the chirp, but the focal geometry places a lower bound on the duration of the peak, and the bandwidth of the pulse limits the range of high intensity to a small fraction of the focal length. An alternative  technique employs an axiparabola \cite{smartsev2019axiparabola} to focus different radial locations in the near field to different axial locations in the far field and an echelon to adjust the relative timing of the radial locations. Along with velocity control, the features of this technique---an extended focal range and a near transform-limited intensity peak in the far field---have enabled a novel regime of laser-wakefield acceleration that eliminates dephasing and greatly decreases the accelerator length \cite{palastro2020dephasingless, caizergues2020phase}. Nevertheless, each velocity requires a unique axiparabola-echelon pair, hindering its tunability.

Each of these methods uses linear optical elements in the near field to structure a pulse with advantageous space-time correlations, but nonlinear processes, such as self-focusing, can also give rise to these correlations \cite{marburger1975self,shen1975self,couairon2007femtosecond}. Self-focusing occurs when the nonlinear optical response of a medium, quantified by the nonlinear refractive index ($n_2$), reduces the phase velocity in regions of high intensity. The ratio of the instantaneous pulse power, $P(t)$, to the critical power, $P_c = {\eta}\lambda _0 ^2/4\pi n_0 n_2$, parameterizes the effect, where $\lambda_0$ is the central, vacuum wavelength of the pulse, $\eta$ depends on its transverse profile, and $n_0$ is the linear refractive index \cite{marburger1975self}. For temporal slices within a pulse with $P(t)>P_c$, self-focusing overcomes diffraction. These slices undergo transverse collapse until their intensity reaches a threshold for activating a mechanism that can arrest the collapse before it reaches a singularity. Because the distance over which a slice collapses depends on its value of $P(t)/P_c$, the temporal profile of the power correlates time within the pulse to a collapse location. This correlation, as previously described by the moving focus model, results in a collapse trajectory for a traditional Gaussian pulse as shown in Fig. 1(a) \cite{shen1975self, couairon2007femtosecond, brodeur1997moving}. 

Here we describe the first nonlinear technique for spatiotemporal control: the "self-flying focus," which combines temporal pulse shaping with the moving focus model to control the trajectory of an intensity peak over distances comparable to the focal length [Fig. 1(b)]. Specifically, the instantaneous power determines the collapse location for each temporal slice, with the minimum and maximum powers setting the collapse range, while the pulse shape determines the time at which the intensity peak moves through these locations. A self-focusing arrest mechanism with an intensity threshold, such as ionization refraction, ensures that the maximum intensity of the peak remains nearly constant throughout the collapse range. Unlike linear techniques, which use specially designed optics to manipulate the space-time structure in the near field, the self-flying focus mixes temporal shaping in the near field with spatial shaping through nonlinear self-focusing in the far field. As an example of its utility, simulations demonstrate that a self-flying focus pulse with an intensity peak that counter-propagates with respect to the group velocity can create a highly uniform, meter-long plasma channel---a critical component of advanced laser and beam-driven accelerators.
\begin{figure}
\centering\includegraphics[width=\textwidth]{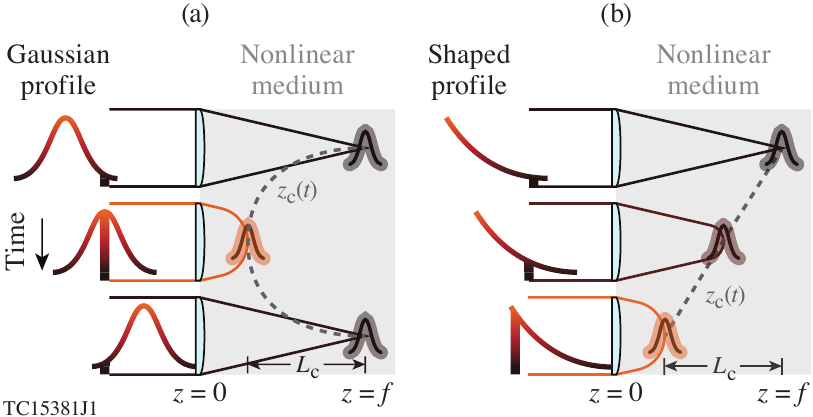}
\caption{Focusing a laser pulse with a Gaussian temporal profile with $P(t)>P_c$ into a nonlinear medium creates a collapse point that traces out a U-shaped trajectory $z_c(t)$ over a distance $L_c$ at a velocity that is decoupled from the group velocity. This is described by the standard moving focus model. (b) Focusing a temporally shaped pulse with $P(t)>P_c$ can create a collapse point and intensity peak that instead follows any arbitrary trajectory---in this case a constant, backward velocity---over a distance $L_c$.}
\end{figure}

\section{Collapse dynamics and the self-flying focus}
Figure 1 illustrates that the trajectory of self-focusing collapse can be controlled with temporal pulse shaping. As described by the moving focus model \cite{shen1975self, couairon2007femtosecond, brodeur1997moving}, a Gaussian pulse with $P(t)>P_c$ focused into a nonlinear medium by an ideal lens exhibits a U-shaped collapse trajectory over the collapse range ($L_c$) [Fig. 1(a)]. The lower-power temporal slices collapse closer to the linear focal point ($z=f$), while the higher-power slices collapse closer to the lens ($z=0$). When temporal pulse shaping is combined with the moving focus model, the collapse point can move at an arbitrary, constant velocity through the collapse region [Fig. 1(b)]. In this example, the power increases with time, ensuring that the lower-power slices collapse \emph{earlier} and further from the lens, while the higher-power temporal slices collapse \emph{later} and closer to the lens. In both cases, the collapse velocity is decoupled from the group velocity of the pulse. For the remainder of this work, we focus on the novel situation shown in Fig. 1(b).

While the constant velocity collapse trajectory shown in Fig. 1(b) could have utility for a number of applications, the self-flying focus has much more flexibility. In fact, a power profile can be found for any desired collapse trajectory. In the source dependent expansion method (SDE) \cite{sprangle2002propagation}, the collapse point as function of power is given by $z_c=[1/f+(P/P_c-1)^{1/2}\lambda_0/\pi n_0 w_i^2]^{-1}$, where $w_i$ is the spot size incident on the lens. Re-expressing this equation in terms of the power provides
\begin{equation}
\frac{P(t)}{P_c}=\left[\left(\frac{w_i}{w_f} \right)\left(\frac{1}{z_c(t)/f}-1 \right) \right]^2 +1,
\end{equation}
where $w_f$ is the linear focal spot. Here $t$ represents time within the pulse. This is distinct from the time at which that temporal slice collapses, i.e., $t_c = t + z_c(t)/v_g$, where $v_g$ is the group velocity. While an arbitrary, desired trajectory, $z_c(t_c)$, can be found by iterating to solve for $z_c(t)$, a constant velocity trajectory has the relatively simple solution. Differentiating $t_c$ with respect to $z_c$, rearranging terms, and integrating yields $z_c(t) = z_c(0) + ut$, where $u = v_gv_c/(v_g-v_c)$ is the reduced velocity and $v_c$ the desired velocity of the collapse point. For the remainder of this paper, $n_0 \approx 1$ and $v_g \approx c$. Note that the profile depends only on the ratio of the initial spot size at the lens to the linear vacuum spot ($w_i/w_f$) and the collapse velocity. A positive $u$ corresponds to strictly positive, subluminal collapse velocities, while a negative $u$ can indicate either a negative collapse velocity or a positive, superluminal collapse velocity. In practice, however, collapse trajectories with positive $u$ values may be impractical, because temporal slices later in the pulse will experience the refractive conditions created by slices earlier in the pulse.

The trajectory persists over a range defined by the minimum and maximum collapse points, $L_c = z_c(P_{min})-z_c(P_{max})$, or, more explicitly,
\begin{equation}
\frac{L_c}{f}=\frac{1}{1+\left(\frac{w_f}{w_i}\right)\sqrt{P_{min}/P_c-1}}-\frac{1}{1+\left(\frac{w_f}{w_i}\right)\sqrt{P_{max}/P_c-1}}.
\end{equation}
Notably, this range can be tuned through the focal geometry or power and can approach distances comparable to the focal length. Scaling $P_{min}$ and $P_{max}$ in tandem shifts the range along the propagation axis, while adjusting the difference between $P_{min}$ and $P_{max}$ alters the total distance. Alternatively, one can alter the distance through $w_i/w_f$. Decreasing $w_i$ decreases the distance it takes a temporal slice to collapse radially, while increasing $w_f$ corresponds to weaker linear focusing, which enhances the relative effect of self-focusing. 

Figure 2 displays power profiles for a constant collapse velocity and illustrates two properties of a self-flying focus pulse. First, the ratio $w_i/w_f$ determines the steepness of the power curve for fixed values of $L_c/f$ and the pulse duration. In effect, a focal geometry with a smaller value of $w_i/w_f$ lowers the maximum power requirement for the pulse. Second, the overall pulse duration (neglecting rise and fall times) is simply $t_p = L_c/|u|$. More generally, a pulse defined by a particular time interval in Fig. 2 will have an associated collapse range $\Delta L_c/f$. As an example, a pulse spanning the interval $\Delta \tau$ = 0.6 - 0.2 will have a collapse range of $\Delta L_c/f = 0.4$, where $\tau\equiv |u| t/f$ is the normalized time.

\begin{figure}
\centering\includegraphics[width=8.4cm]{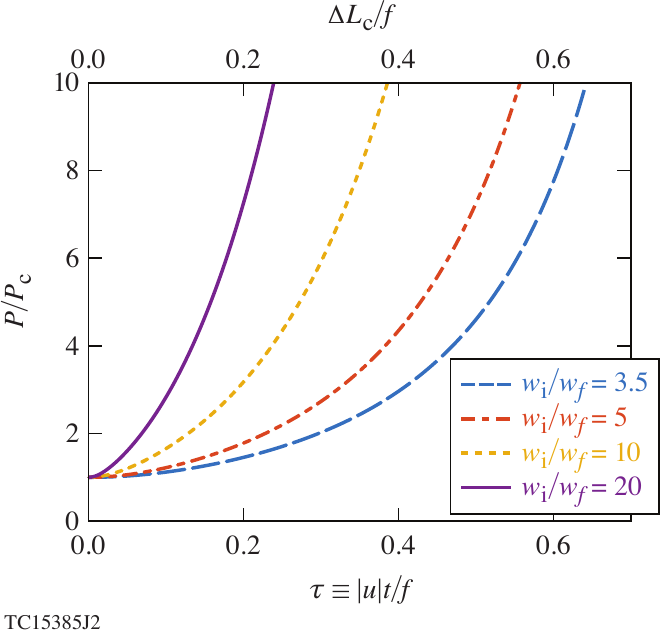}
\caption{Power profiles for constant velocity collapse trajectories with $u < 0$, or upon reversing the horizontal axis, $u > 0$. For fixed $\Delta L_c/f$, the ratio $w_i/w_f$ sets the steepness of the profile. Here, the minimum power is set to $P_c$, but the profiles for different values of $P_{min}$ [and corresponding $z_c(0)$] can be found by shifting each curve to the left.}
\end{figure}

Depending on the desired parameters, the duration of a self-flying focus pulse can be quite large (e.g., $t_p = 2$ ns for $v_c = -c$, $L_c/f = 0.3$, and $f = 1$ m). However, the effective duration of the intensity peak formed by the collapse of adjacent time slices within the pulse can be much shorter. In media where the self-focusing arrest mechanism has an intensity threshold $(I_a)$, (e.g., due to ionization, harmonic generation, or absorption \cite{couairon2007femtosecond}), the effective duration will scale as $t_e \propto w_a^2/v_c\lambda_0$, where $w_a \approx \sqrt{2P/\pi I_a}$ is the spot size at arrest. That is, the effective duration of the intensity peak is proportional to the difference in time over which adjacent slices self-focus and refract. While arrest always occurs before collapse, the trajectory of the resulting intensity peak will nearly match that of the collapse point as long as $w_i >> w_a$. This property---that the duration of the intensity peak can be substantially shorter than the pulse duration---opens up the possibility of using long pulses for applications that typically require short pulses. As a result, the self-flying focus could take advantage of existing long pulse, high energy laser systems with advanced pulse shaping capabilities, such as at the National Ignition Facility or the OMEGA laser.

\begin{figure}
\centering\includegraphics[width=8.4cm]{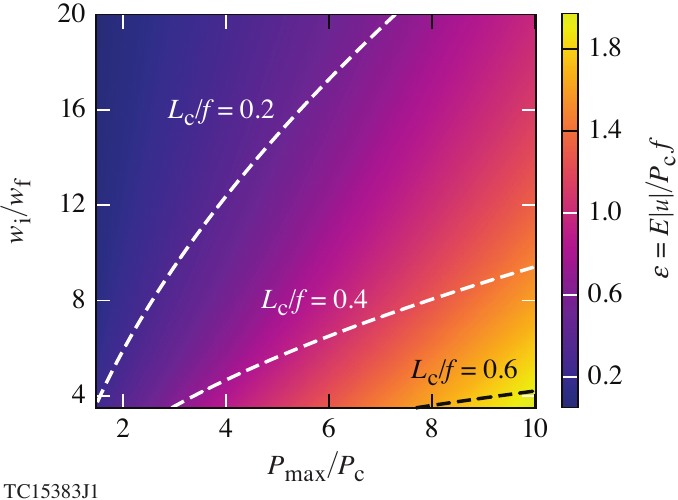}
\caption{The energy design space for a self-flying focus pulse as a function of the focal geometry, peak power of the laser system, and desired collapse range (dashed contours). Along an $L_c/f$ contour, the energy requirement grows with increasing spot ratio and power. A larger maximum power provides a larger spot at arrest ($w_a$). The initial power is set to $P_c$ ($z_0=1$) for simplicity.}
\end{figure}

The energy expenditure for a self-flying focus pulse can be tuned to meet the requirements of a broad range of laser systems (Fig. 3). Integrating the power in Eq. (1) over time and defining the normalized energy as $\varepsilon \equiv  |u|E/P_cf$ provides
\begin{equation}
\varepsilon = \frac{L_c}{f} + \left(\frac{w_i}{w_f} \right)^2\left\{\frac{L_c}{f}\left[1+z_0^{-1}\left(z_0 \pm \frac{L_c}{f}\right)^{-1}\right] + 2\ln\left[ z_0^{\pm 1}\left(z_0\pm \frac{L_c}{f}\right)^{\mp 1}\right]\right\},
\end{equation}
where $z_0\equiv z_c(0)/f$ and the $\pm$ accounts for positive and negative reduced velocities ($u$), respectively. For a particular lens (i.e., a focal length) and a desired $L_c/f$, the required energy can be adjusted through $w_i/w_f$. Along an $L_c/f$ contour, the energy requirement grows with increasing spot ratio and peak power. Larger energies correlate with larger maximum powers, which, in turn, will result in larger spot sizes at arrest. As also indicated by Fig. 2, for fixed $w_i/w_f$, a larger maximum power lengthens the collapse range ($L_c/f$). However, this has a potential downside of greater variation in the spot size at arrest: $\sqrt{2P_{min}/\pi I_a}$ to $\sqrt{2P_{max}/\pi I_a}$. The scaling $E=\varepsilon P_cf/|u|$ exhibits the expected behavior that a self-flying focus can be created with less energy in media with a stronger nonlinearity (i.e., a smaller critical power). Of particular note is that the required pulse energy decreases with increasing $|u|$ due to the shorter pulse duration, $t_p = L_c/|u|$. This property could be useful for applications in which precise velocity matching is unnecessary, but controlling the direction of $v_c$ is, such as the formation of plasma channels or in directed energy.

The equations presented for power, collapse range, and energy all depend on the collapse distance ($z_c$). No exact expression exists for this distance for an arbitrary transverse profile. The SDE method used in Eqs. (1)-(3) assumes that the transverse profile remains Gaussian until collapse \cite{sprangle2002propagation}; physically, a beam undergoing collapse will evolve to the characteristic Townes profile \cite{moll2003self}. While the simplicity is convenient for developing theoretical scalings and illustrating the salient physics, more precise predictions require a collapse distance determined by simulations or measurements. In general, the self-focusing collapse distance of a focused temporal slice can be expressed as
\begin{equation}
z_c = \left\{ \frac{\lambda_0 \sqrt{[(P/P_c)^{1/2}-\beta]^2-\gamma}}{\alpha\pi n_0 w_i^2} + \frac{1}{f} \right\} ^{-1},
\end{equation}
where $\alpha$, $\beta$, and $\gamma$ are curve-fitting parameters that depend on the transverse profile. The SDE method predicts $\alpha = 1$, $\beta = 0$, and $\gamma = 1$, but a more accurate simulation analysis of a Gaussian profile conducted by Marburger found $\alpha = 0.367$, $\beta = 0.852$, and $\gamma = 0.0219$ \cite{marburger1975self}. In all figures and simulations presented here, the Marburger values were used. 

In arriving at Eqs.(1)-(3), it was assumed that $f\gtrapprox 3Z_R$ ($w_i/w_f \gtrapprox3.2$), where $Z_R\equiv \pi w_f^2n_0/\lambda_0$ is the linear Rayleigh range of a focused slice. Further, the maximum power was limited to $\approx10P_c$ to minimize the likelihood of transverse breakup and altered scalings prior to collapse. For pulses with reasonably low levels of phase and intensity noise, this limit should be conservative and could potentially be pushed to $\approx40P_c$ \cite{fibich2005self}. As a final note, multiple self-focusing/refocusing cycles could extend the region of high intensity beyond $L_c$ and the lens focal length \cite{kosareva2011arrest}.  

\section{Simulation and demonstration}
The ability to control the intensity trajectory over long distances makes the self-flying focus ideal for creating long plasma channels---a critical component in a number of applications, such as advanced laser-based accelerators and directed energy. Current techniques for creating long plasmas rely on filamentation through a dynamic balancing of self-focusing and plasma refraction \cite{braun1995self,couairon2007femtosecond}, axicon focusing \cite{durfee1993light,akturk2009generation, green2014laser}, variable wave front distortion \cite{ionin2013filamentation}, or the use of short wavelengths \cite{shipilo2017fifteen}. Axicon focusing, for example, can suffer from significant pump depletion and ionization refraction by the end of the medium due to the forward propagation of the intensity peak \cite{akturk2009generation}. The self-flying focus has elements in common with filamentation, but offers velocity control and does not necessarily require a short-pulse laser. Further, the ability to counter-propagate the intensity peak with respect to the pulse avoids ionization refraction, allowing for a wider range of focal geometries \cite{turnbull2018ionization}.

Here the self-flying focus is applied to the formation of a plasma channel necessary for the recently described ``dephasingless'' laser-wakefield accelerator \cite{palastro2020dephasingless, caizergues2020phase}. The simulations solve the paraxial wave equation assuming cylindrical symmetry with source terms accounting for self-focusing, plasma refraction, and loss due to field ionization and inverse bremsstrahlung (c.f., Ref. \cite{palastro2018ionization}). The gas is initially neutral and the electron density evolves due to multiphoton and collisional ionization alongside radiative and three-body recombination.  

Figure 4(a) displays results for a self-flying focus pulse with $v_c =-c$ propagating through Lithium gas and triggering a sharp ionization front that travels at that same velocity (i.e., $v_c =-c$) over a meter. The negative collapse velocity allows the intensity peak to propagate through the background gas, rather than the ionized plasma, mitigating ionization refraction. The specific velocity of $v_c=-c$ was chosen such that an injected, relativistic electron bunch would be velocity matched to the plasma creation and thus experience constant plasma conditions throughout its acceleration. 

Figures 4(b) and (c) demonstrate that subluminal backwards ($v_c=-c/2$) and superluminal forwards ($v_c=2c$) velocities can also create long plasma channels. When $v_c=2c$, the shorter effective duration ($t_e$) results in less initial plasma heating, allowing recombination to set in. Nevertheless, re-self-focusing of the higher power temporal slices leads to eventual, full ionization. 

\begin{figure}
\centering\includegraphics[width=\textwidth]{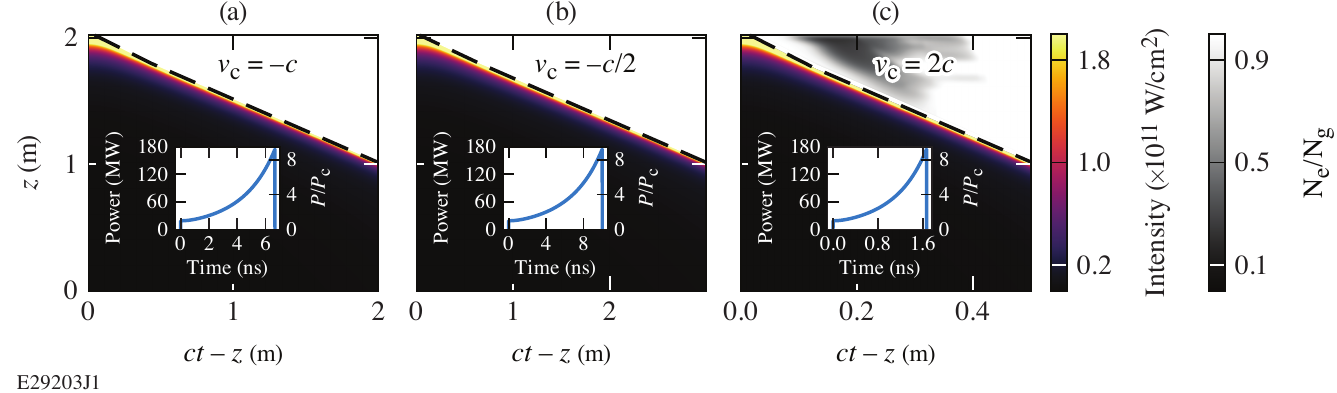}
\caption{On-axis intensity and electron density profiles from simulations of a $\lambda_0 = 1 \mu$m self-flying focus pulse with three different collapse velocities, $v_c = -c$ (a), $v_c = -c/2$ (b), and $v_c = +2c$ (c) propagating through a lithium gas of density $N_g = 10^{19}\textrm{cm}^{-3}$. The incident power profile is plotted in each inset, with 20 ps exponential rise and fall times added to better represent a realistic shaped pulse. For these three collapse trajectories, the intensity peak resulting from the collapse moves exactly at the desired trajectory and creates a smooth ionization front along that trajectory.}
\end{figure}

Figure 5 illustrates the additional self-focusing cycles for $v_c = -c$.  The leading edge of the intensity peak advances forward in the moving frame ($ct-z$) as temporal slices earlier in the pulse approach their collapse location.   Refraction from the near-fully ionized plasma arrests the collapse at a spot size $w_a \approx 100\mu m$. After their initial refraction from the plasma, the later temporal slices undergo several more self-focusing cycles.

The density of the lithium gas, $N_g=10^{19}\textrm{cm}^{-3}$, was chosen to optimize the energy gain of electrons in a dephasingless laser wakefield accelerator: in the relatively compact, 1m plasma created here, electrons could be accelerated to $\approx$ 200 GeV. Lithium, in particular, was selected for its low ionization energy and critical power, $P_c\approx 19\textrm{MW}$ at $\lambda_0 = 1.054 \mu\textrm{m}$ and the density of interest \cite{miles1973optical}. On account of the low critical power, the self-flying focus pulse only required $\approx400$ mJ of energy with a maximum power of $\approx180$ MW, which are well within the capability of existing laser systems. In fact, the laser and lens parameters, an f/500 lens, spot ratio $w_i/w_f\approx 6$, and collapse range $L_c/f = 0.5$ ($f=2$ m, $L_c=1$ m), were chosen to ensure full ionization of the valence state while minimizing the pulse energy. For these parameters, simulations attempted with $0<v_c \le c$ (including standard Gaussian pulses) exhibited such violent and rapid refraction from the generated plasma that the transverse wavenumbers and temporal variations were invalid in the paraxial approximation.

Throughout the collapse range, the maximum on-axis intensity remains nearly constant (within a factor of $2$ of the maximum) despite the power varying by nearly a factor of 10.  Note that because $L_c/f < 1$ a length of gas ahead of the collapse range is required for nonlinear focusing. As a result, not all of the gas is ionized. In principle, this can be overcome by using an additional Kerr lens in the near field to pre-focus each temporal slice before it enters the nonlinear medium. Regardless, the simulation illustrates that the self-flying focus can produce long uniform plasma channels with modest power and energy requirements. 
\begin{figure}
\centering\includegraphics[width=\textwidth]{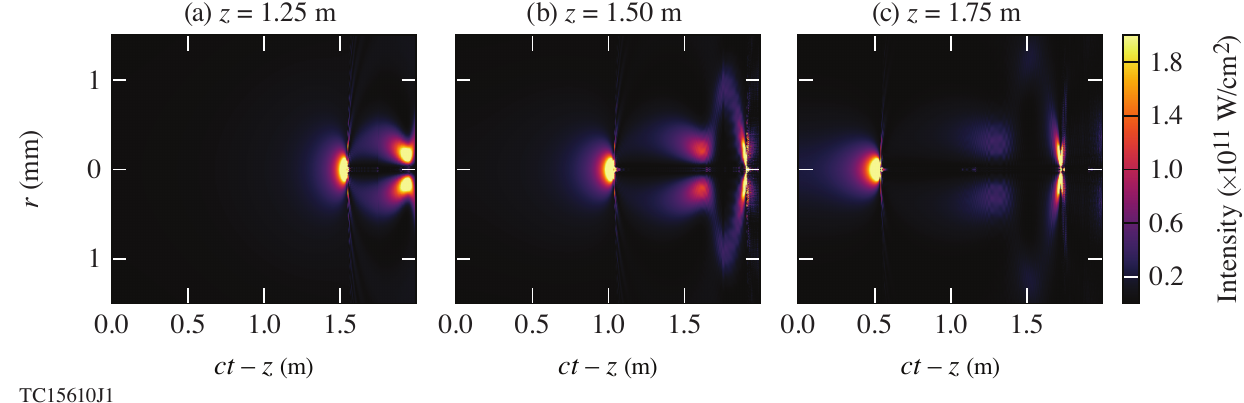}
\caption{Radial intensity profiles of a $v_c = -c$ self-flying focus pulse at three propagation distances ($z = 1.25\textrm{m}$, $z = 1.50\textrm{m}$, and $z = 1.75\textrm{m}$) as a function of the moving frame coordinate ($ct-z$). The leftmost intensity peak at each location corresponds to the intensity peak resulting from the initial collapse, while the intensity peaks at later times result from re-self-focusing. The leading edge of the peak advances forward in the moving frame unhindered by upstream plasma formation. }
\end{figure}

The parameters of the self-flying focus can be readily tuned for applications that require lower-density plasmas ($N_e<10^{17}\textrm{cm}^{-3}$), such as electron or proton beam-driven wakefield accelerators \cite{joshi2018plasma,litos2014high,turner2019experimental}. Here the design would use a gas with a lower $P_c$, (e.g. rubidium or cesium) and, ideally, a shorter wavelength laser (e.g., $\lambda_0 = 351$nm) to offset the higher $P_c$ associated with the lower gas density. While the high intensities associated with the collapse and arrest dynamics of the self-flying focus are suitable for these plasma applications, they may be undesirable in situations where ionization and material damage are unwanted, e.g., THz generation in a crystal. 

\section{Summary and conclusions}
A novel technique for nonlinear spatiotemporal control delivers an arbitrary trajectory intensity peak over distances comparable to the linear focal length. The technique combines temporal pulse shaping in the near field with nonlinear focusing in the far field to control the time and location at which each temporal slice within a pulse undergoes self-focusing collapse/arrest. In principle, the collapse point of these self-flying focus pulses can travel in either the forward or backward direction. In practice, however, only backwards traveling or superluminally forwards travelling collapse points can avoid the refractive conditions created by earlier time slices within the pulse, making the intensity peak trajectory robust and preferable for applications. The self-flying focus can accommodate a wide range of parameters facilitating their use on various laser systems and in diverse applications. Notably, the self-flying focus could take advantage of long-pulse, high-energy laser systems, such as the National Ignition Facility or the OMEGA laser, to create intensity peaks with durations comparable to short-pulse lasers. Simulations demonstrate that a self-flying focus pulse can create a meter-scale plasma channel with a leading edge that is velocity matched to relativistic electrons, promising to both enable and improve this necessary component of advanced accelerators. Further, this demonstration suggests that the self-flying focus could improve the formation of long, uniform plasma channels in other media for filamentation and directed energy-based applications.   

\section*{Acknowledgments}
The authors would like to thank P. Tzeferacos, J.L. Shaw, M. Virgil Ambat, and K.L. Nguyen for fruitful discussions. The work published here was supported by the US Department of Energy Office of Fusion Energy Sciences under contract no. DE-SC0016253, the US Department of Energy Office of Science, Office of High Energy Physics (HEP) Accelerator Stewardship program under Award Number DE-SC0020396, the Department of Energy under cooperative agreement no. DE-NA0003856,  the University of Rochester, and the New York State Energy Research and Development Authority. 

This report was prepared as an account of work sponsored by an agency of the U.S. Government. Neither the U.S. Government nor any agency thereof, nor any of their employees, makes any warranty, express or implied, or assumes any legal liability or responsibility for the accuracy, completeness, or usefulness of any information, apparatus, product, or process disclosed, or represents that its use would not infringe privately owned rights. Reference herein to any specific commercial product, process, or service by trade name, trademark, manufacturer, or otherwise does not necessarily constitute or imply its endorsement, recommendation, or favoring by the U.S. Government or any agency thereof.

\section*{Disclosures}
The authors declare no conflicts of interest.

\bibliography{SelfFF.bib}
\end{document}